\documentstyle[11pt,newpasp,twoside,epsf]{article}
\def\edcomment#1{\iffalse\marginpar{\raggedright\sl#1\/}\else\relax\fi}
\reversemarginpar

\begin{document}
\title{Some Effects of Galaxy Collisions in a Cluster ICM} 
 \author{Curtis Struck \& Jason R. Brown}
\affil{Dept. of Physics \& Astronomy, Iowa State Univ., Ames, IA
50011  USA}
%\author{Co-authors}
%\affil{Their affiliation}

\begin{abstract}
We present preliminary results of numerical modeling of the combined
effects of ICM stripping and galaxy collisions on galaxies in
clusters. We use the Hydra adaptive mesh, P3M + SPH code of Couchman
et al. (1995). Generally, the formation of extended tidal and splash
structures in galaxy interactions will facilitate ram pressure
stripping. ICM compression of massive clouds stripped from tails and
plumes may produce significant populations of free-floating star
clusters or dwarf galaxies.

\end{abstract}

There has been suggestions in the literature that galaxy collisions
are more common in the environment of galaxy clusters (e.g., Fried
1988, Rubin, Waterman \& Kenney 1999). The cumulative effects of long
range tidal disturbances, i.e., galaxy harrassment, can produce some
similar results (see Moore et al. 1996), and so complicates the
interpretation of observation. However, secular processes are unlikely
to have had time to affect gas-rich galaxies in groups or subclusters
falling into large clusters. In this case cluster potential may induce
interactions.

There are some beautiful candidate examples of galaxy collisions in
clusters. These include: 1) NGC 4438/35 = Arp 120 studied by Kenney et
al. (1995, and Kenney \& Yale, 2002). These authors also find evidence
of an AGN or starburst driven outflow. 2) NGC 4485/90 studied by
Clemens et al. (2000, with refs. to earlier papers). Following the
encounter, the smaller galaxy may have had all of its gas stripped. 3)
NGC 4388 studied by Yoshida et al. (2002, and poster at this meeting),
which has long, extra-disk, emission line spirals like those in the
models below.

We would expect that the gas in tidal tails and splash bridges or
plumes would be more easy to strip than gas in disks, and because of
this galaxy collisions would enhance ram pressure stripping. We also
expect ICM compression to facilitate star formation in gas clouds
contained within tidal structures, while ram pressure may push some of
these clouds free of their parent galaxies.

To date there has been little modeling work undertaken to study these
possibilities. We are beginning work in this area, and present
preliminary results here. Specifically, we initialize a galaxy
consisting of an isothermal, spherically symmetric dark matter halo,
plus disk consisting of half star and half gas particles distributed
symmetrically around the x-y plane, The galaxy is impacted by an hot
ICM 'wind' coming from the positive z-direction, and travelling
parallel to the z-axis with a velocity of 2000 km/s relative to the
galaxy disk, as in the models of Schulz and Struck (2001). In
addition, we define a circular region within the disk, centered at
half the disk radius (x=0.5, y=0.5 in model units), with its radius
equal to about one third of the galaxy disk radius. Within that circle
the gas particles are given a negative velocity in the z direction of
300 km/s, as though the disk had been directly impacted by a smaller
disk.

The impacted gas is pushed down, but not immediately, nor entirely
stripped (see Figure 1). Eventually, we find that stripping is
enhanced by this collisional disturbance - about 30\% of the disk gas
is removed, versus 18\% in the corresponding non-collisional
case. This is true even without the formation of extended tidal tails,
which should be easy to strip. The remnant disk is much more disturbed
in the collisional case.

As in stripping calculations without a collision, ram pressure
stimulates the formation of spiral arms in the gas disk (via the
annealing process of Schulz and Struck 2001). The outer parts of these
arms are stripped and compressed, with the formation of clumps in the
process. These clumps are good candidate sites for the formation of
young star clusters or dwarf galaxies. We would expect the same
phenomena in tidal tails. 

Angular momentum transfer is more vigorous in the collision than in
the comparison run. As a result, gas in the remnant disk core is more
compressed and would likely be the site of a subsequent starburst.

\begin{figure}[h]
%\plotone{fig1.eps}
\plotfiddle{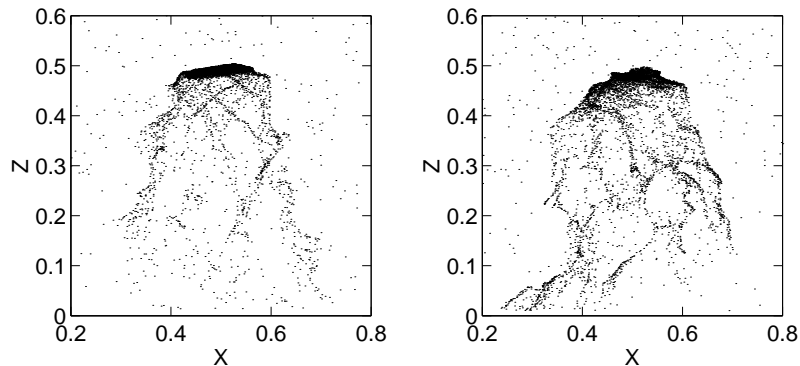}{1.55in}{0}{60}{60}{-175}{-175}
\caption{Edge-on views of gas particles in a galaxy disk moving
face-on into an ICM wind at a time of 94 Myr. after the onset of the
wind. The grid scale is 100 kpc. The model in the right panel has
collided with a smaller disk as described in the text.} 
\end{figure}

\end{document}